# Systematic Literature Reviews in Software Engineering - Enhancement of the Study Selection Process using Cohen's Kappa Statistic

Jorge Pérez, Jessica Díaz, Javier Garcia-Martin, Bernardo Tabuenca
jorgeenrique.perez, yesica.diaz, javier.garciam, bernardo.tabuenca @upm.es
Universidad Politécnica de Madrid, 28031 Madrid, Spain

**Abstract.**

**Context**: Systematic literature reviews (SLRs) rely on a rigorous and auditable methodology for minimizing biases and ensuring reliability. A common kind of bias arises when selecting studies using a set of inclusion/exclusion criteria. This bias can be decreased through dual revision, which makes the selection process more time-consuming and remains prone to generating bias depending on how each researcher interprets the inclusion/exclusion criteria.

**Objective**: To reduce the bias and time spent in the study selection process, this paper presents a process for selecting studies based on the use of Cohen's Kappa statistic. We have defined an iterative process based on the use of this statistic during which the criteria are refined until obtain almost perfect agreement (k>0.8). At this point, the two researchers interpret the selection criteria in the same way, and thus, the bias is reduced. Starting from this agreement, dual review can be eliminated; consequently, the time spent is drastically shortened.

**Method**: The feasibility of this iterative process for selecting studies is demonstrated through a tertiary study in the area of software engineering on works that were published from 2005 to 2018.

**Results**: The time saved in the study selection process was 28% (for 152 studies) and if the number of studies is sufficiently large, the time saved tend asymptotically to 50%.

**Conclusions**: Researchers and students may take advantage of this iterative process for selecting studies when conducting SLRs to reduce bias in the interpretation of inclusion and exclusion criteria. It is especially useful for research with few resources.

## 1. Introduction

Evidence-based software engineering (EBSE) aims at providing *"the means by which current best evidence from research can be integrated with practical experience and human values in the decision making process regarding the development and maintenance of software"* [17]. Systematic literature review (SLR) is an important methodology of the EBSE paradigm and has various objectives [16], [6], [8]: (i) to summarize the existing evidence concerning a practice or technology, (ii) to identify gaps in current research, (iii) to help position new research activities, and (iv) to examine the extent to which a hypothesis is supported or contradicted by the available empirical evidence [9]. SLRs identify, evaluate, and interpret all available relevant research on a specified research question or topic area using a rigorous and auditable methodology [19]. According to Dingsøyr & Dybå [13], a key feature that distinguishes systematic reviews from traditional narrative reviews is that the former make an *"explicit attempt to minimize the chances of drawing wrong or misleading conclusions as a result of biases in primary studies or from biases arising from the review process itself"*. Zhang & Ali Babar [31] argue that SLR processes and protocols should be rigorously described to minimize biases, which can be prevalent in traditional reviews, and ensure the reliability of reviews and their reproducibility under the same conditions. Kitchenham & Charters [19] define three main phases of an SLR: (i) *planning the review*, in which a review protocol is developed; (ii) *conducting the review*, in which the protocol that was planned in the previous phase is executed; and (iii) *reporting the review*, in which the review steps are presented to the community. The review protocol defines the methods for undertaking a systematic review, thereby reducing the possibility that the review can be influenced by research expectations (bias). The review protocol must specify the search strategy of the studies; the selection criteria, namely, the inclusion and exclusion criteria (IC/EC) to be applied during the selection of primary studies; the data extraction method; and the synthesis strategy.

Few SLRs quantify the necessary effort for performing these reviews; e.g., [3]. Zhang and A. Babar concluded that the most time-consuming activities were study selection and data extraction [31]. The reliability and repeatability of the study selection process depend strongly on the degree of bias when performing this process [30]. We focus on *selection bias* [4], which is generated when the selection criteria are not sufficiently clear or contain ambiguities. Additional factors that bias the selection of primary studies, such as conflicts of interest of authors and sponsors, are not considered in this work. However, the more unbiased, auditable and repeatable the SLR methodology is, the more effort and time are required.

Peer review (dual revision) is the most common method for reducing the bias during the selection of primary studies. However, this method lengthens the selection process and is still prone to generating bias, the severity of which depends on how each researcher interprets and applies the IC/EC. To reduce the bias and time spent in the study selection process, this paper presents an enhancement of the selection process that is based on the use of Cohen's Kappa statistic to measure the level of agreement between the inter-raters, namely, the two researchers who are responsible for selecting the studies. The use of Cohen's Kappa statistic is a more robust approach than the observed proportion of agreement because Kappa considers the effect of chance. We propose using Cohen's Kappa statistic to measure the level of agreement regarding the application of the IC/EC in an iterative process for selecting studies. During the iterative process, the criteria are refined until an almost perfect agreement is reached. At this point, the two researchers understand and interpret the IC/EC in the same way and, thus, the bias is reduced. Then, the dual revision process stops, and the two researchers can apply the selection criteria individually on the remaining studies. Thus, a substantial part of the effort that is devoted to the study selection process is saved.

Therefore, this work defines an iterative process for selecting studies and deal with Cohen's Kappa interpretation, including the first paradox of the statistic, to reduce the selection bias and eliminate work overload during the study selection process. We have defined an iterative process for refining the IC/EC toward avoiding dual review based on the Kappa values. The feasibility of this iterative process for selecting studies is evaluated in a tertiary study in which publications are reviewed in 4 scientific databases, 13 journals and 3 conferences in the area of software engineering from January 2005 to July 2018. Also, this tertiary study will show that the process that we proposed has not been applied previously in software engineering research.

The remainder of the paper is structured as follows: Section 2 describes the background. Section 3 presents the main contribution of this work: an improved study selection process for SLRs. Section 4 presents a tertiary study as a case study for which we describe how to realize the proposed enhancements in the selection process. Finally, the conclusions and limitations of the study are described in Section 5.

## 2. Background

This section describes (i) the study selection process of an SLR and the bias problem that is generated by the reviewers' interpretations of the IC/EC and (ii) Cohen's Kappa statistic, which is used to measure the level of interrater agreement, and its first paradox.

## 2.1 Study Selection Process & Bias Problem

The study selection process involves both reviewing the studies that are identified in the search and selecting the studies that are relevant to the objective of the SLR against the previously defined IC/EC. A lack of bias when performing this selection process is necessary for ensuring the reliability and repeatability of the process [31]. A bias can be defined as *"a systematic error, or deviation from the truth, in results or inferences"* [13]. According to Kitchenham et al. [21], the SLR methodology aims at being as unbiased as possible by being auditable and repeatable.

Typically, the selection of primary studies is a two-stage process [7]. First, at least two researchers review the titles and abstracts of studies that are identified by the search and irrelevant papers are rejected (preprocessing). According to Brereton et al. [7], if researchers cannot agree, the paper should be included. Second, full copies of the papers are reviewed by at least two researchers against the IC/EC. The two researchers should resolve any disagreements with the help of an independent arbitrator, if necessary.

Thus, the primary studies are selected according to an interpretation of the previously established IC/EC. Selection bias [4] can be generated when the selection process is driven by research expectations or the selection criteria are not sufficiently clear or contain ambiguities. McDonagh, Peterson, Raina & Chang [25] states that *"even when reviewers have a common understanding of the selection criteria, random error or mistakes may result from individual errors in reading and reviewing studies"*.

The bias problem can be mitigated whenever two researchers perform the selection process. According to Budgen et al. [10], all decisions about the IC/EC should be based on an analysis by two of the reviewers, working in various pairings to help minimize bias. Similarly, McDonagh, Peterson, Raina & Chang [25] considers dual review to be sufficient for ensuring the reliability of the study. Finally, Zhang & Ali Babar [31] conclude that peer review is the most common method for reducing bias, as it is used by 80% of systematic reviewers. However, dual review implies that two reviewers evaluate all articles; thus, the required effort and time increase considerably. More important, this method might generate biases that depend on how each researcher applies the IC/EC; hence, it is difficult to guarantee the reliability and repeatability of the selection process.

Zhang & Ali Babar [31], via a survey of 52 respondents, discovered that reviewers used additional methods to complement dual revision, such as external checkers, self-review, and validation of agreements via statistical techniques, e.g., Kappa. Hence, Kitchenham & Brereton [18] suggest that whenever there might be discrepancies on whether to include a study or not, both reviewers should discuss it until an agreement is reached. Da Silva et al. [12] also recommend that studies should be selected by at least two researchers. The results of the selection are integrated into an Agreement / Disagreement Table (ADT), which is evaluated by a third researcher; disagreements are discussed among all researchers and resolved by consensus. This task is more time-consuming as an additional researcher is involved. Moreover, it does not take "agreements by chance" into account.

An alternative procedure for mitigating the bias problem without incurring a considerable increase in time is to have only one reviewer evaluate all articles and a second reviewer evaluate only those articles that were excluded by the first reviewer. Typically, this alternative saves effort but still generates bias. Indeed, the second reviewer might be influenced in reviewing the studies by the knowledge that the first reviewer excluded them [25]. Another alternative procedure is to revise only 10% to 20% of the studies, to refine the description of the IC/EC [25]. However, there is no evidence that 10-20% dual revision is sufficient for concluding that the IC/EC are correctly interpreted by the two reviewers when these criteria are applied individually on the complete set of studies. It is necessary to measure the level of agreement between the inter-raters to validate the sufficiency of that pilot test.

Kitchenham & Charters [19] proposed using Cohen's Kappa statistic to measure the agreement between the two researchers that assess each paper during the study selection process. However, they did not describe how to use it, when to use it, or how many times to use it, nor did they consider the paradoxes of the statistic. Dingsøyr & Dybå [13] also state that decisions about study eligibility are typically made by two independent reviewers to increase the reliability and discuss the convenience of using Cohen's Kappa to measure the agreement between the researchers. However, the authors also did not describe how to use this statistic. Kitchenham & Brereton [18] describes an example in which Kappa values are calculated in several phases of the SLR process, such as selection validation and data extraction. These Kappa values are calculated after both researchers have performed the dual review.

Ali & Petersen [2] proposed the following procedure: Two researchers apply the "think-aloud protocol" over five randomly selected studies—namely, the two reviewers express out loud the reasons that lead them to include or exclude a study. This is intended to clarify ambiguities and misinterpretations of the selection criteria and contributes to improving the internal consistency of studies. Then, both reviewers apply the IC/EC to a common random subset of selected studies (a pilot investigation of 20 studies), the interrater agreement is calculated, and the disagreements are discussed. Here, Cohen's Kappa statistic and the observed proportion of agreement are used to evaluate the level of agreement. Finally, all primary studies are reviewed in a dual way. Currently, Kappa is the most frequently recommended coefficient for measuring interrater agreement [13][18]. Ali & Petersen [2] use both the Kappa value and the observed agreement value. This strategy is aligned with [26] that state that "*Cohen's kappa is a more robust method than percent agreement since it is an adjusted agreement considering the effect of chance*". However, the Kappa value can be misleading. If the Kappa value is too low and the observed agreement is high, we are faced with the first paradox of the statistic. The paradox occurs because Kappa is sensitive to the distribution of the data. Therefore, it is desirable to present both (1) the observed proportion of an agreement and (2) the Kappa coefficient. Park & Kim [26] propose using alternative statistics. However, we think is not necessary to use alternative statistics. There are means for interpreting the value of Kappa, which will be described later.

**2.2 First Paradox of Cohen's Kappa Statistic**

Cohen's Kappa coefficient measures the concordance between two judges' classifications of N elements into C mutually exclusive categories. Cohen defined the coefficient, which is denoted as k, as "*the proportion of chance-expected disagreements which do not occur, or alternatively, it is the proportion of agreement after chance agreement is removed from consideration*" [11]. The coefficient is calculated via the following formula:

$$k = \frac{p_0 - p_c}{1 - p_c} \qquad (1)$$

$p_0$ = the proportion of units for which the judges agreed (relative observed agreement among raters)
$p_c$ = the proportion of units for which agreement is expected by chance (chance-expected agreement)

Table I presents the contingency matrix, which specifies the frequency distributions of the categories for the two judges. From Table I, $p_o$ and $p_c$ are calculated as follows:

$$p_o = P(a) + P(d) \quad (1.1)$$
$$p_c = P_{category1} + P_{category2} \quad (1.2)$$
$$P_{category1} = (P(a) + P(c)) * (P(a) + P(b)) \quad (1.2.1)$$
$$P_{category2} = (P(b) + P(d)) * (P(c) + P(d)) \quad (1.2.2)$$

TABLE I
CONTINGENCY MATRIX

|  |  | Judge 1 | |
|---|---|---|---|
|  |  | *Category 1* | *Category 2* |
| **Judge 2** | *Category 1* | a: number of agreements on category 1<br>P(a) = a/N | b: number of disagreements (judge 1 and category 2, and judge 2 and category 1)<br>P(b) = b/N |
|  | *Category 2* | c: number of disagreements (judge 1 and category 1, and judge 2 and category 2)<br>P(c) = c/N | d: number of agreements on category 2<br>P(d) = d/N |

The coefficient k=0 when agreement equals chance agreement. Greater-than-chance agreement corresponds to a positive value of k and less-than-chance agreement corresponds to a negative value of k. The maximum value of k is +1.00, which occurs when (and only when) there is perfect agreement between the judges [11]. Landis & Koch [23] proposed the following table for evaluating intermediate values (Table II).

TABLE II
INTERPRETATION OF K VALUES (FROM [23])

| Kappa Statistic | Strength of Agreement |
|---|---|
| < 0,00 | Poor |
| 0.00 – 0.20 | Slight |
| 0.21 – 0.40 | Fair |
| 0.41 – 0.60 | Moderate |
| 0.61 – 0.80 | Substantial |
| 0.81 – 1.00 | Almost Perfect |

Under various conditions, the k statistic is affected by two paradoxes that return biased estimates of the statistic itself. Nonetheless, to the best of our knowledge, many researchers do not consider these paradoxes when they interpret the coefficient. We focus on the first paradox as it affects more directly the interpretation of the Kappa value of the interrater agreement during the study selection process. Feinstein & Cicchetti [14] emphasized the influence of the first paradox: "*The first paradox of k is that if $p_c$ is large, the correction process can convert a relatively high value of $p_0$ into a relatively low value of k*". This conversion is caused by a substantial imbalance in the marginal totals (horizontal and vertical) of the contingency matrix. This imbalance is generated by the prevalence of one trait (category) versus the other. In general, for the same proportion of observed agreements, the closer to 0.5 the prevalence is (the more balanced the marginal totals are), the greater the value of Kappa is. In other words, very low or very high prevalence of one category penalizes the Kappa coefficient because, in that case, the proportion of agreements that are expected by chance ($p_c$) is greater compared to cases in which prevalence is close to 0.5.

When Cohen's Kappa is applied to the study selection process in SLRs, the judges, i.e., the reviewers, classify each study into two categories: included or excluded. When high agreement is observed, a large proportion of the studies may have been included (or excluded). In this case, the "included" trait has a high prevalence compared to the other trait and, consequently, a low value of Kappa would be obtained. As an example, two authors of this work have recently conducted an SLR in the domain of microservices [1]. Both researchers carried out the process of selecting the primary studies and defined the IC/EC. First, a master's student performed the search of primary studies in a set of prefixed databases, journals and conferences. Second, the student delivered a common set of 10 primary studies to the two researchers, who included or excluded the studies according to the established selection criteria. Table III presents the 2x2 contingency matrix for the review of the 10 studies. The obtained values are as follows: $p_0 = 0.8$, $p_c = 0.68$, and $k = 0.375$.

TABLE III
CONTINGENCY MATRIX EXAMPLE

|  |  | Judge A | |
|---|---|---|---|
|  |  | Yes | No |
| Judge B | Yes | 1 | 1 |
|  | No | 1 | 7 |

The relative observed agreement between the raters is high (80%), whereas the Kappa value is low (0.375). This is caused by the substantial imbalance in the marginal totals (horizontal and vertical) of Table III. This imbalance is caused by the prevalence of "no" versus "yes". Lantz & Nebenzahl [24] posit that this problem can be resolved by defining a sample of balanced prevalence at the outset as an element of the experimental design. An alternative solution is to explore Kappa by calculating the maximum ($k_{max}$), the minimum ($k_{min}$) and the normal value ($k_{nor}$). Independent of the relative observed agreement between raters for a specified value of $p_0$, the $k_{min}$ and $k_{max}$ values represent a range of possible values for k. These variables are calculated as follows [24]:

$$k_{max} = \frac{p_0^2}{(1-p_0)^2+1} \qquad (2)$$

$$k_{min} = \frac{p_0-1}{p_0+1}; \text{ for } p_0 < 1 \qquad (3)$$

$$k_{nor} = 2p_0 - 1 \qquad (4)$$

Significant deviations of k from $k_{nor}$ in any direction suggest the existence of predominant asymmetry in the agreement category or, alternatively, in the disagreement category. In the example above-mentioned, $K_{max}= 0.57$, $k_{min}= -0.25$ and $k_{nor}= 0.6$. A significant deviation of the value of k (0.375) from the $k_{nor}$ value can be observed, which indicates a predominant asymmetry, in this case, of the agreement categories. Lantz & Nebenzahl [24] recommend that the k value be reported together with the $p_0$, $S_D$ and $P^{++}$ values (or $S_A$ and $P^{--}$ values), where $S_D$ is an asymmetry index for the disagreement prevalence, $P^{++}$ represents the agreement prevalence for one of the traits, $S_A$ is an asymmetry index for the agreement prevalence, and $P^{--}$ represents the disagreement prevalence for one of the traits.

$$S_D = \frac{P(b)-P(c)}{1-p_0} \qquad (5) \qquad P^{++} = P(a) \qquad (6)$$

$$S_A = \frac{P(a)-P(d)}{p_0} \qquad (7) \qquad P^{--} = P(d) \qquad (8)$$

## 3. Enhancements of the Study Selection Process Using Cohen's Kappa Statistic

Study selection processes are typically carried out by two researchers [10], [25], [31]. As discussed in Section 2, dual revision is the most common method for minimizing bias. First, both researchers review all the studies that were identified in the search stage. Then, they exclude or include these studies based on the previously defined IC/EC. Finally, the classifications are compared between the two researchers (see Fig. 1).

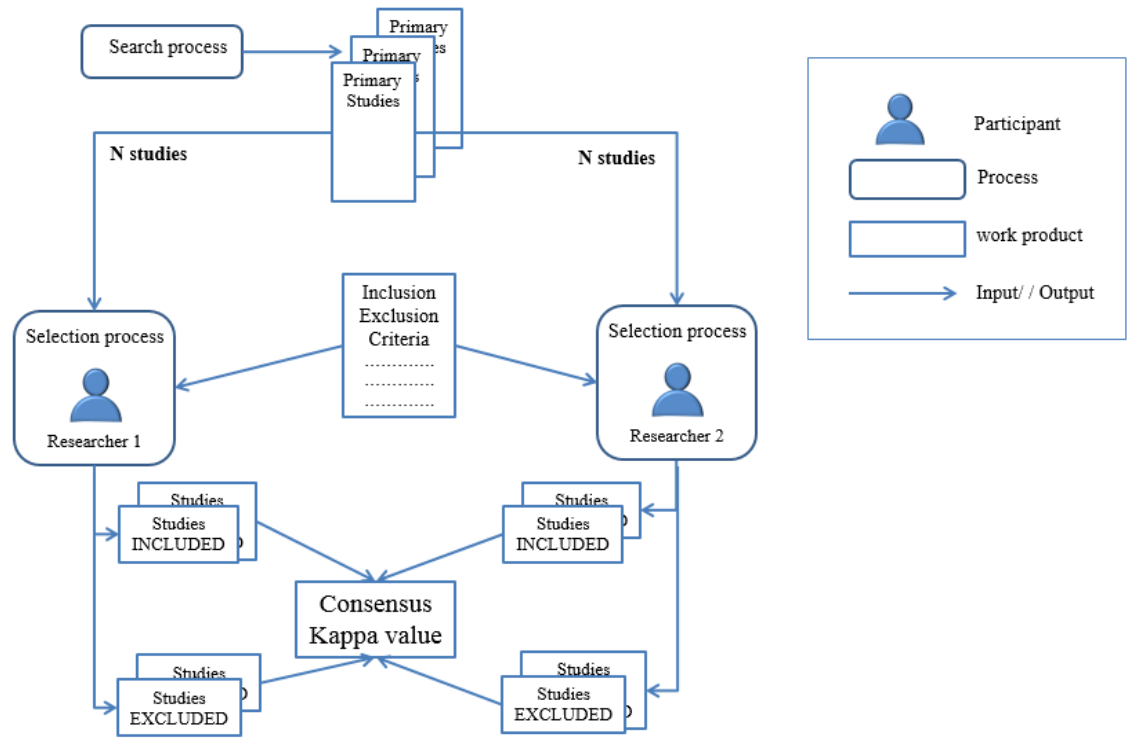

Fig. 1. Traditional peer-review study selection process

Discrepancies are usually resolved by consensus after discussing both interpretations of the IC/EC [18]. Alternatively, discrepancies are resolved by the judgement of a third researcher [12]. Cohen's Kappa is sometimes used to explore the level of agreement between the reviewers [2], [13], [18]. However, this measure is typically used to report the final agreement between the reviewers when all the studies have been analyzed; consequently, it does not modify or improve the selection process. Hence, Cohen's Kappa statistic is not used to mitigate work overload during the study selection process. It is not an iterative process for refining the IC/EC toward avoiding dual review based on the Kappa values, but rather a measure of overall agreement between the two researchers on the primary studies that are analyzed. In this proposal we could have used other statistics to measure the level of agreement between the inter-raters such as Fleiss's K [15] or Krippendorff's Alpha [22], but the community of software engineering recommends the use of Cohen's Kappa.

This section presents a study selection process that improves upon the process that was initially described by [19]. The selection process (see Fig. 2.1) presupposes, such as the previous one, the existence of an established IC/EC and a dual revision. It also assumes the existence of N studies that are obtained from the previous search process. This improved process consists of two phases, which are described as follows.

**Phase 1**. The first phase consists of a dual revision that is iteratively performed. In each iteration, a set of 15 studies are randomly selected from the N studies, and then they are revised by both reviewers. The value 15 is arbitrary; you can select any other. However, from our experience, a set of 15 studies is a sufficiently large sample for identifying differences in the interpretation of the selection criteria between the reviewers; and 15 studies do not represent a large percentage of the total population. Next, two reviewers analyze the titles, abstracts, and keywords and, if necessary, the section of conclusions of the common set of 15 studies. The reviewers annotate their decisions to include or exclude the studies, together with the IC or EC that led the judgement. Finally, the

Kappa value (k) is calculated. If k≤0.8, the reviewers should discuss those studies on which they have disagreed to clarify how they have applied the IC/EC. As a result of this phase, the selection criteria are refined, and the interpretation of new criteria should be less ambiguous and more consensual. Phase 1 is repeated until k>0.8. A new set of 15 studies is selected and reviewed. If k>0.8, it is assumed that the reviewers applied the IC/EC in a consistent manner and phase 2 begins.

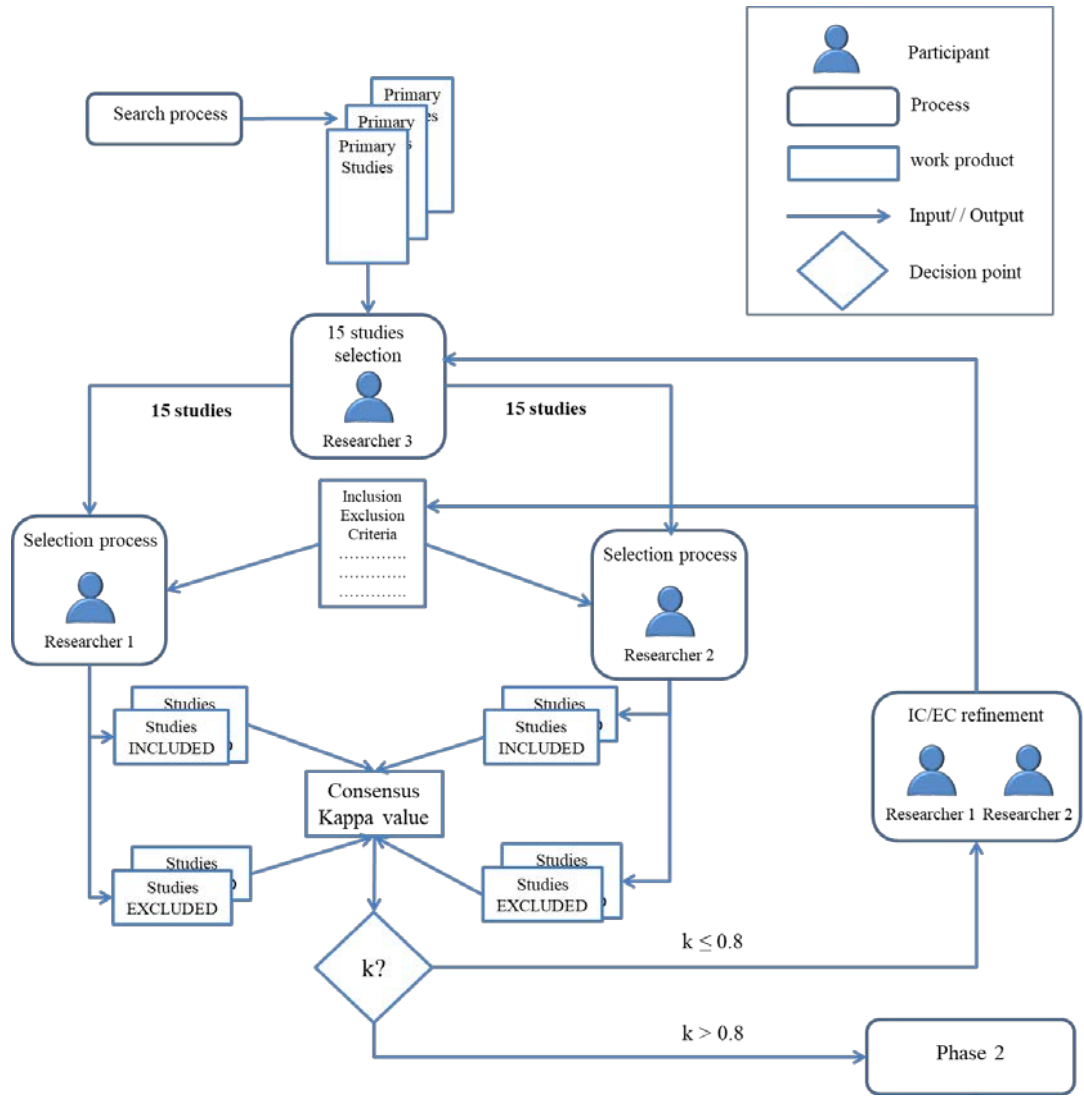

Fig. 2.1. Enhancement of the study selection process (Phase I)

**Phase 2**. In this phase each study is analyzed by a single reviewer. The two reviewers individually apply the selection criteria—iteratively refined in Phase 1—to the studies that remained after applying Phase 1, which are divided in two sets—one set for each reviewer (see Fig. 2.2).

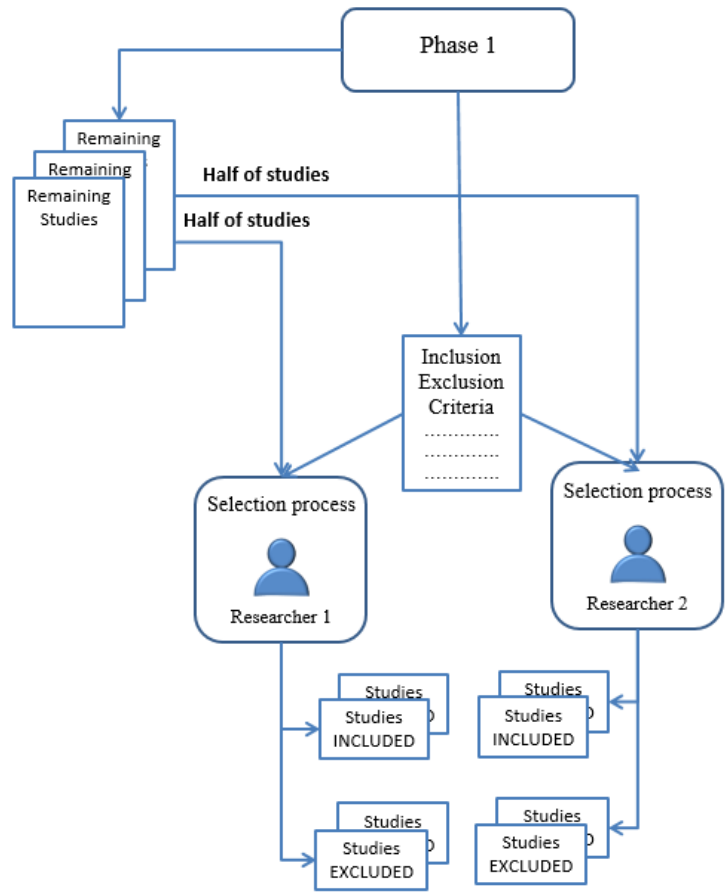

Fig. 2.2. Enhancement of the study selection process (Phase II)

In our experience, k>0.8 is found after the second or third round [27]. It is necessary to assess both the Kappa value and the observed proportion of agreement. If the observed proportion of agreement is high and the Kappa value is low, we are faced with the first paradox of the statistic. In this case, the coefficients that are described in Section 2.2— see equations (2), (3), and (4)—should be calculated to place the interrater agreement into perspective.

This procedure is an iterative process for realizing a higher agreement level between the researchers that perform the selection process. The expression of selection criteria is refined until both researchers agree on how to apply them. This procedure has two key advantages: 1) Time is saved when the agreement between the researchers has been reached as, from that moment, dual review is reduced to single review. 2) Bias is reduced as several rounds of reviews are executed to guarantee that both researchers agree on the selection criteria. The value of k indicates when to stop these rounds.

This same process can also be applied to the preparation of Systematic Mapping (SM). In these, the criteria are broader and therefore more rounds will be needed (we assume).

## 4. Case study: A Tertiary Study

This section aims to provide empirical evidence that validates that the use of the iterative process for selecting studies is feasible. Inter-rater agreements and time savings are calculated by applying the proposed process. Next, the case study is reported according to the guidelines for conducting and reporting case study research in software engineering by Runeson and Höst [28]. The goal of reporting a case study is twofold: to communicate the findings of a study, and to work as a source of information for judging the quality of the study. With this twofold goal, the reporting of the case study is described as follows.

The feasibility of the iterative process for selecting studies is evaluated via a tertiary literature review of SLRs published between January 2005 and July 2018 in the software engineering domain. The objective of this tertiary study is to demonstrate the use of the enhancements in the study selection process described in Section 3. In addition, this study corroborates that the proposed process had not been applied previously in software engineering research by analyzing how SLRs on software engineering from 2005 to 2018 perform the study selection process.

**4.1 Case Study Design**

This section describes the case study, i.e., the research questions that are the focus of this case study, the subjects participating in the case study, as well as data collection, analysis, and validation procedures.

4.1.1 Research objective and questions

Evidence of the feasibility of process was obtained by putting the process into practice in a real SLR, which in this case was a tertiary study. The research questions to be answered through the case study analysis can be formulated as follows: **RQ1**: Does the proposed procedure minimize the bias during the study selection process in a systematic review? **RQ2**: Does the procedure imply a reduction in the time devoted by the researchers to perform the study selection process?

4.1.2 Subject description

This case study involved 4 researchers (R1 to R4) that contributed to conduct the tertiary study. R1, R2, R3 and R4 were respectively the authors of this paper in the same order they are presented in the title page. Briefly, R1-R4 were involved in the search process of the SLR, R1-R3 were involved in the selection process using the iterative process here described, and R1 and R3 were involved in the extraction process.

4.1.3 Case study description & Data collection procedure

The case study consisted of conducing a tertiary study. Therefore, the case study description consisted of the description of the review plan. To that end we followed the guidelines Kitchenham & Chartes [19] as follows: In the review planning phase, a review protocol was developed that specified (i) the review objective and research questions; (ii) the search process; (iii) study selection process; (iv) the data extraction strategy; and (v) the strategy for synthesizing the extracted data. As the case study aimed at demonstrating the feasibility and time savings of our proposal for the selection process of studies, we mainly focused on the steps (i), (ii) and (iii).

Research question: **[SLR]RQ**: Is the study selection process here described a novel process?

Search Process: A formal search strategy was required for identifying the entire population of scientific papers that could be relevant to the objective of this study. The search strategy defines the search space: electronic databases, journals and conference proceedings were considered key spaces for the review objective (see Table IV). During the SLR search phase authors had to expend a lot of time and overcome a large number of barriers [5]. To obtain the journals and conference proceedings that were essential for the objective of this tertiary study, we reviewed a set of outstanding published articles on the topic of SLRs, which mostly focused on software engineering [10], [21], [18],

[20], [31], [12], [29]. Regarding electronic databases, we thought that the use of the 4 indicated in Table IV were sufficient guarantee to cover the entire spectrum of publications made in the domain of software engineering.

We established that two researchers (R2 and R3) had to perform an automatic search in electronic databases using the following query string: ("systematic literature review" OR "systematic review" OR "systematic mapping study" OR "mapping study" OR "literature review" OR "literature survey" OR "meta-analysis) AND ("software engineering"). We established to filter by the title, abstract and keywords of the articles. Additionally, two researchers (R1 and R4) had to perform a manual search in a set of conference proceedings and journals (see Table IV), in which they filtered by title.

TABLE IV
SEARCH SOURCES

| Data Source | Documentation |
| --- | --- |
| Electronic Databases | Scopus<br>WOS (Web of Science)<br>IEEE (Xplore Digital Library)<br>Science Direct |
| Conference proceeding manual searches | Evaluation and Assessment in Software Engineering.<br>Empirical Software Engineering and Measurement.<br>International Conference on Software Engineering. |
| Journal manual searches | ACM Computing Surveys<br>Advanced Engineering Informatics<br>Communications of the ACM<br>Computer<br>Empirical Software Engineering<br>IEEE Transactions on Software Engineering<br>IEEE Software<br>IET Software<br>Information & Software Technology<br>Journal of Software-Evolution and Process<br>Journal of Systems and Software<br>Software – Practice & Experience<br>ACM Transactions on Software Engineering and Methodology |

Study Selection Process: This section describes the protocol for selecting the studies that were relevant to the review objectives according to a set of IC/EC. The study selection process was the same as we described in Section 3. The review protocol also specified IC and EC, which determined whether each study should be considered or not for this systematic review (see Table V). A researcher (R3) selected 15 studies and two researchers (R1 and R2) independently analyzed the full texts of these 15 studies, determined whether each study was included or excluded, and filled in Table VI. In this particular case, it was necessary to analyze the full text to answer the inclusion criteria number 2. Both researchers met to contrast their results, refine the IC/EC (if applicable), and calculate and comment on the values of k, $S_D$ and $P^{++}$. The data to be collected in each iteration are listed in Table VII.

TABLE V
SELECTION CRITERIA

**Inclusion criteria**
1. Studies (SLRs, SMSs, literature surveys, or meta-analyses) that are written in English according to the research string pattern that is defined in the protocol;
2. Studies that have a well-defined description of the primary study selection process;
3. Studies that are within the software engineering domain.

**Exclusion criteria**
1. Studies that are outside the software engineering domain;
2. Studies that deal with approaches/tools for improving/automating SLRs, SMSs, literature surveys or meta-analysis studies;
3. Studies (SLRs, SMSs, literature surveys, or meta-analyses) that focus on processes other than the selection of primary studies;
4. Studies (SLRs, SMSs, literature surveys, or meta-analyses) that are based on a methodology that lacks a primary selection process;
5. Papers for which only PowerPoint presentations or extended abstracts were available;
6. Short papers (less than 6 pages).

TABLE VI.
STUDY SELECTION FORM TEMPLATE

| Inclusion criteria (of the current iteration) | | Exclusion criteria (of the current iteration) | |
|---|---|---|---|
| Reviewer: | | | |
| Study ID | Study title | Include? (Y/N) | IC/EC |
| | | … | … |
| | | | Time spent (hh:mm): |

TABLE VII.
DUAL STUDY SELECTION SUMMARY FORM TEMPLATE

| Study ID | Study title | Include? (Y/N) | | IC/EC (comments) | |
|---|---|---|---|---|---|
| | | R1 | R2 | | |
| … | … | … | … | …. | |
| k= | $k_{max}=$ | $k_{min}=$ | | $k_{nor}=$ | $S_D=$ | $P^{++}=$ |
| Comments | | | | | |

This process was repeated while $k \leq 0.8$. When $k>0.8$, the dual review stopped, and each researcher received half of the remaining studies, which were randomly selected by R2, to complete the study selection process. All time that was spent on this process was annotated by the researchers who were involved.

4.1.4 Analysis and validity procedure

In this case study, both qualitative and quantitative data were gathered. Qualitative data were collected from the iterative process during which IC/EC were refined. Quantitative data were collected from the values of k, $k_{nor}$, $k_{min}$, $k_{max}$, and time taken during the iterative process of selection of studies. To increase the validity of the case study, observer triangulation was applied. Two external reviewers (see acknowledgements) were involved in the case study by replicating specific data collection sessions by these two different observers. All collected data—included the references of these papers and results of the process—are available in a public repository (https://drive.upm.es/index.php/s/emaAEmItedvswEb) to motivate others to provide similar evidence by replicating this tertiary study.

## 4.2 Results

This section describes the analysis and interpretation of the results of conducting the SLR, as well as the evaluation of its validity.

### 4.2.1 Case study execution

This section describes the conduction and reporting of the SLR, specifically the results for the study searching and selection process, and results of the SLR research question.

Search: Following the review protocol described in Section 4.1, a search for secondary studies was carried out. The search strings that were used in each electronic database are specified in the "*search strings*" file of the repository. We located 2,438 studies from search resources that were defined in the protocol (see Fig. 3), of which 1,080 were duplicates. Additionally, it was not possible to obtain the full-text articles for 80 studies. Thus, the systematic review retrieved 1278 unduplicated scientific papers, which are listed in the "*search results*" file of the repository.

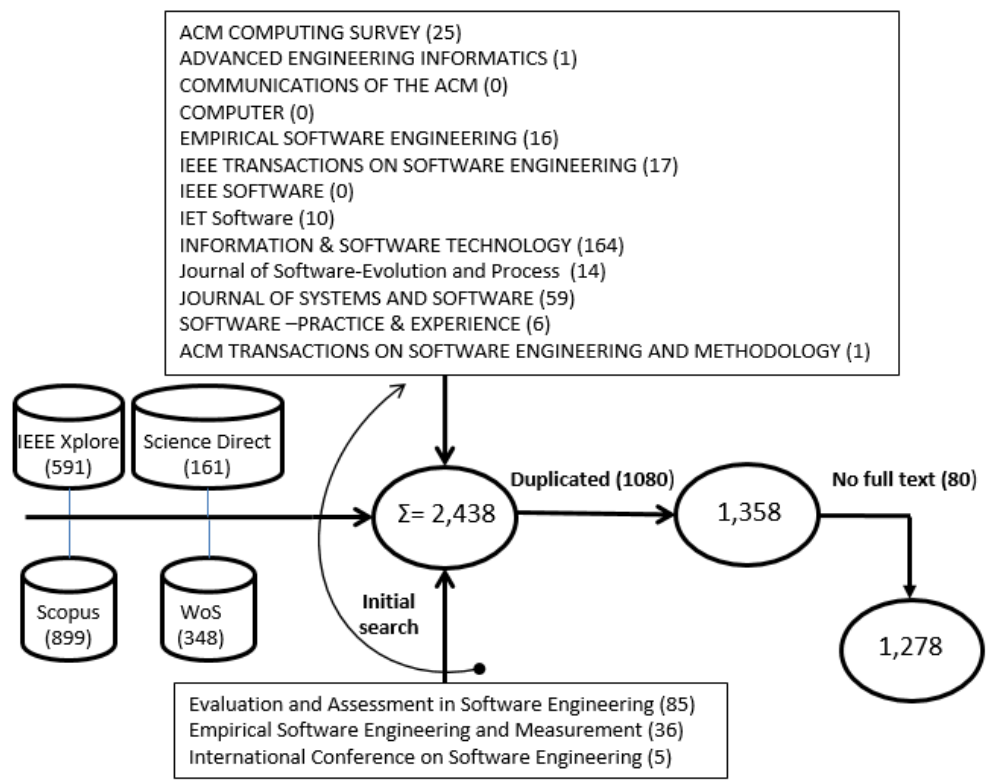

Fig. 3. Search process results

Selection: As 1278 secondary studies is too many for the objective of the case study, we randomly selected 152 secondary studies. This size of sample (152) is sufficiently large for measuring the time savings and the interrater agreement, but also small enough to be affordable as a case study. The process of selecting studies implied 3 iterations in **phase 1**. In **iteration 1**, R3 assigned the same 15 studies to researchers R1 and R2. Both applied the IC/EC previously defined in the review protocol (see Table V). The results were stored in the "*R1 01-15*" and "*R2 01-15*" files of the repository. Table VIII (iteration 1) presents the contingency matrix and the Kappa values that were obtained from the analysis. The value of k (0.7) did not exceed 0.8; therefore, phase 1 was repeated. R1 and R2 discussed the discrepancies during the application of IC/EC to refine the criteria (See Table VIII). These are the results concluded in the discussion:

- R1 included study ID 35705, whereas R2 did not because it was a master's thesis. As this criterion had not been defined, a new inclusion criterion, namely, "*Studies that are published in conference/workshop proceedings, journals, and book chapters,*" and a new exclusion criterion, namely, "*Studies such as theses, editorials, and books that were not subjected to a standardized peer-review process*" were added. Finally, study ID 35705 was excluded.
- R1 did not include study ID 35260, while R2 did. The problem was the interpretation of inclusion criterion 2 and exclusion criterion 3: the concept "primary study selection process" needed to be defined more precisely. Finally, study ID 35260 was included.
- R1 excluded studies ID 20381, ID 5023, ID 5040 and ID 35552. R2 also excluded these studies, but the criterion for exclusion was different. There was a problem in the interpretation of exclusion criteria 2 and 3, which needed to be refined.

In **iteration 2**, R3 selected another 15 studies. R3 assigned the same set of 15 studies to researchers R1 and R2. Both applied the IC/EC refined in iteration 1. The results were stored in the "*R1 16-30*" and "*R2 16-30*" files of the repository. Table VIII (iteration 2) presents the contingency matrix and the Kappa values obtained from the analysis. The value of k (0.74) did not exceed 0.8; therefore, phase 1 was repeated. R1 and R2 discussed the discrepancies during the application of the IC/EC to refine the criteria (See Table VIII). These are the results concluded in the discussion:

- R1 excluded study ID 5340 by applying the criterion 1, while R2 included it. Both researchers concluded that it was a literature survey and study ID 5340 was included.
- R1 excluded study ID 4822 by applying the criterion 3, while R2 included it. Again, the disagreement was due to the interpretation of the concept "primary study selection process". The researchers decided to relax this criterion. Finally, study ID 4822 was included.

During **iteration 3**, R3 selected another 15 studies. R3 assigned the third set of 15 studies to researchers R1 and R2. Both applied the IC/EC refined in iteration 2. The results of applying these criteria were stored in the "*R1 31-45*" and "*R2 31-45*" files of the repository. Table VIII (iteration 3) presents the contingency matrix and the Kappa values that were obtained from the analysis of these 15 studies. The value of k (1) exceeds 0.8; therefore, the dual review ends and phase 2 begins.

During **phase 2** reviewers R1 and R2 individually applied the criteria to the 107 studies that remained after phase 1. These studies were split into two sets where R1 analyzed 54 studies, and R2 analyzed 53 studies. The files "*R1 phase2*" and "*R2 phase2*" of the repository list the reasons for the inclusion/exclusion of each study.

In summary, during phase 1, R1 and R2 applied the IC/EC to the same studies (45 studies) until the two researchers apply these criteria in a homogeneous manner (i.e. the value of k exceeded 0.8). During phase 2, R1 and R2 individually applied IC/EC to 107 studies. During phase1, R1 and R2 selected for extraction 31 of the 45 initial studies. During phase 2, R1 selected 34 studies and R3 selected 35 studies. In aggregate, 100 (of the 152 studies that constituted the sample) were selected for the extraction process.

TABLE VIII.
RESULTS SUMMARY OF PHASE 1 (DUAL REVISION)

**Iteration 1** (studies 1-15)

|    |     | R1  |    |
|----|-----|-----|----|
|    |     | Yes | No |
| R3 | Yes | 9   | 1  |
|    | No  | 1   | 4  |

| $k= 0.7$ | $k_{max}= 0.74$ | $k_{min}= -0.07$ |
|---|---|---|
| $k_{nor}= 0.73$ | $S_D= 0$ | $P^{++}= 0.6$ |

Inclusion criteria (after iteration 1)
1. Studies (SLRs, SMSs, literature surveys, or meta-analyses) that are written in English according to the research string pattern that is defined in the protocol.
2. Studies that are published in conference/workshop proceedings, journals, and book chapters.
3. Studies that have a well-defined description of the primary study selection process. This description includes inclusion/exclusion criteria and an explanation of their application to the primary studies.
4. Studies that are within the software engineering domain.

Exclusion criteria (after iteration 1)
1. Studies that do not include an SLR, SMS, literature survey, or meta-analysis. Examples are studies that present approaches/tools for improving/automating SLRs or SMSs.
2. Studies such as theses, editorials, and books that were not subjected to a standardized peer-review process. Papers for which only PowerPoint presentations or extended abstracts were available. Short papers (less than 6 pages).
3. Studies (SLRs, SMSs, literature surveys, or meta-analyses) that do not describe the primary study selection process.
4. Studies that are outside the software engineering domain.

**Iteration 2** (studies 16-30)

|    |     | R1  |    |
|----|-----|-----|----|
|    |     | Yes | No |
| R3 | Yes | 7   | 2  |
|    | No  | 0   | 6  |

| $k= 0.74$ | $k_{max}= 0.74$ | $k_{min}= -0.07$ |
|---|---|---|
| $k_{nor}= 0.73$ | $S_D= 1$ | $P^{++}= 0.47$ |

Inclusion criteria (after iteration 2)
1. Studies (SLRs, SMSs, literature surveys, or meta-analyses) that are written in English according to the research string pattern that is defined in the protocol.
2. Studies that are published in conference/workshop proceedings, journals, and book chapters.
3. Studies that have a well-defined description of the primary study selection process. This description includes at least the inclusion/exclusion criteria.
4. Studies that are within the software engineering domain.

Exclusion criteria (after iteration 2)
1. Studies that do not include an SLR, SMS, literature survey, or meta-analysis. Examples are studies that present approaches/tools for improving/automating SLRs or SMSs.
2. Studies such as theses, editorials, and books that were not subjected to a standardized peer-review process. Papers for which only PowerPoint presentations or extended abstracts were available. Short papers (less than 6 pages).
3. Studies (SLRs, SMSs, literature surveys, or meta-analyses) that do not describe the primary study selection process (that is, they do not include the inclusion/exclusion criteria).
4. Studies that are outside the software engineering domain.

**Iteration 3** (studies 31-45)

|    |     | R1  |    |
|----|-----|-----|----|
|    |     | Yes | No |
| R3 | Yes | 13  | 0  |
|    | No  | 0   | 2  |

| **$k= 1$** | $k_{max}= 1$ | $k_{min}= 0$ |
|---|---|---|
| $k_{nor}= 1$ | $S_D=$ | $P^{++}= 0.87$ |

Results: *[SLR]RQ: Is the process here described a novel process?* We performed an extraction process on the secondary studies selected in the previous step to determine whether the iterative process for selecting studies that we proposed here had been previously used. The studies that refer to the Kappa statistic have been compiled in the file "studies with kappa" of the repository.

The analysis of the results shows that some of the studies found in the review performed iterations to increase the value of Kappa; however, they do not explain whether they are dealing with the first paradox of the statistic. These studies do not typically specify the observed proportion of agreement nor report the interpretation of the Kappa value versus the observed proportion of agreement. A few studies mention the first paradox (indicating that it is described in the literature) when they compare the value of the observed proportion of agreement versus the Kappa value; however, they do not discuss it further.

According to our analysis, 12 of the 100 secondary studies refer to the Kappa statistic (8 of them published between 2015 and 2018). However, although these 12 SLRs use Cohen's Kappa to measure the agreement between the researchers, they do not indicate to what extent the inclusion/exclusion criteria were refined or modified (after the Kappa value was obtained) to reduce the selection bias, nor do these SLRs indicate whether dual revision was eliminated once a specified Kappa value had been attained. Thus, our work involves a more rigorous application of the statistic than the studies that we analyzed. Our goal is not to increase the value of Kappa, but to reduce the selection bias and the time spent on study selection. Cohen's Kappa statistic is only an instrument and not an end.

4.2.2 Analysis and interpretation

The proposed methods helped the researchers to refine the IC/EC even twice (iteration 1 and iteration 2, see Table VIII), and thus, reduce misunderstandings and bias. The amount of time that was spent by each researcher during the selection process was also recorded. Table IX lists these amounts of time, along with the corresponding researcher, task, and phase. The researchers spent 05:25 hours on phase 1 of the selection process, namely, on the three iterations before reaching a value of k≥0.8. Additionally, they spent 05:09 hours on phase 2, namely, on the processing of 107 studies individually (with an average of 2.89 minutes per study). Thus, they spent a total of 10 hours and 44 minutes on processing the 152 studies in the sample.

TABLE IX.
AMOUNTS OF TIME THAT WERE SPENT BY RESEARCHERS DURING THE SELECTION OF PRIMARY STUDIES

| Researcher | Task | Phase | Time (hh:mm) |
|---|---|---|---|
| R3 | Selection of 45 studies to be delivered to R1 and R3 | 1 | 00:30 |
| R1 | Selection study process against IC/EC (01-45) | 1 | 02:20 |
| R2 | Selection study process against IC/EC (01-45) | 1 | 02:05 |
| R1 & R2 | Meetings (2) for discussion selection criteria | 1 | 00:30 |
| R3 | Selection of 54 and 53 studies to be delivered to R1 and R3, respectively | 2 | 00:10 |
| R1 | Selection study process against IC/EC | 2 | 02:24 |
| R2 | Selection study process against IC/EC | 2 | 02:45 |
| | | Σ | **10:44** |

From this data, we can conclude that the time spent on the study selection process was reduced applying the iterative process described in Section 3. We estimated that two reviewers might invest 07:19 hours each (a total of 14:38) to process 152 studies using the traditional study selection process. Using our iterative process, we spent 10:44. Thus, the time savings is 28%. Consequently, the savings produced by the iterative process for selecting studies described in Section 3, tend asymptotically to 50% as the number of studies to be reviewed increases (see appendix A).

4.2.3 Evaluation of validity

Construct validity is concerned with the procedure to collect data and with obtaining the right measures for the variables to being studies. This threat was mitigated through observer triangulation. However, the main limitation in case study research is external validity, i.e. the generality of results. A limitation of our work is that the proposed iterative process for selecting studies has only been put into practice in this work and two additional works of two of the authors of this work [1][27]. Also, to determine whether the iterative process for selecting studies that we proposed here had been previously used, it has only been carried out on the 100 studies selected for extraction and not on the 1278 studies results of the search process. Thus, further replications and analysis are necessary for generalizing the conclusions. To mitigate this external validity, we hardly worked on the replicability of the study. This study provides detailed information on how the tertiary study was conducted to facilitate the reproduction of the study. Additionally, the complete datasets and descriptive documents necessary to repeat our study are available in a public repository.

**4.3. Case Study Conclusions**

We obtained evidence of the feasibility of the proposed process for selecting studies in SLRs as well as the time saving during this phase of SLRs. To that end we conducted of a case study that consists of a tertiary study. The results show evidence of that (1) the process reduced misunderstandings and bias in the interpretation of IC/EC, and (2) the time taken by reviewers was reduced.

**6. Conclusions**

This work defined an iterative process for selecting studies. It utilizes Cohen's Kappa to reduce the selection bias and the amount of time that is devoted to selecting studies and considers the first paradox of the statistic. The feasibility of this iterative process was demonstrated in a tertiary study in software engineering, specifically on papers that were published from 2005 to 2018. This tertiary study demonstrated the viability of using Kappa to decrease the bias and time spent on the primary study selection process. Section 4.2.1 described an iterative process of refining the IC/EC until the two researchers apply these criteria in a homogeneous manner ($k > 0.8$), thereby reducing the selection bias. The value of the observed proportion of agreement and the possible appearance of the first paradox of the statistic were considered. It was also demonstrated that if the number of studies to be processed is sufficiently high, the savings approach 50% asymptotically.

Although the use of Kappa is recommended during the primary study selection process for reducing the selection bias, in practice, it is used infrequently. Only 12% of the studies that were analyzed in this research refer to the statistic. Furthermore, these works use the Kappa value only to obtain the final value: the interrater agreement is calculated after the complete set of studies has been processed. To interpret the value of Kappa, the value of the observed proportion of agreement must be calculated and the possible appearance of the first paradox of the statistic and how to solve it must be considered. This last issue is only briefly mentioned in a few studies. A few pilot studies have been conducted to refine the inclusion/exclusion criteria; none suppress dual review above a specified Kappa value. The analyzed pilot studies do not explain whether, in the event of the paradox, they have calculated the $k_{max}$, $k_{min}$ and $k_{nor}$ values to place Kappa into perspective.

The iterative process that is presented in this paper will help researchers and students on software engineering who perform SLRs improve the study selection in two ways: by avoiding selection bias and reducing the amount of time that is spent on it. The proposed use of Kappa to enhance and improve SLR is an important research methodology for EBSE.

As future work, the proposed method can be updated to include quality gates during the individual review to validate if inter-rater agreement is still valid. This means that, once *k>0,8* and individual review starts, after X studies the value of *k* is calculated for a set of papers in dual review. Finally, it would be interesting to check if, once IC/EC have been refined (*k>0,8*), the rest of selection process could be conducted by N judges who had not participated in this criteria refinement, as considered by Fleiss's K [15] or Krippendorff's Alpha [22].

### Acknowledgements

This study was performed with the assistance of Andrea Villegas and Gladys Martín, who actively participated in the main study and the tertiary study for demonstrating the use of the proposed approach.

## Appendix-A: Time saving

During phase 1 of the proposed process there is a dual revision for a time $T_0$. From $T_0$, both reviewers are put to work in different set of studies (i.e. each study is analyzed by a single reviewer). Suppose v the average velocity to review studies per instant of time. After the training time $T_0$, the reviewers analyzed $S_0 = vT_0$ studies. Now, suppose it is necessary to analyze S studies, being $S \geq S_0$. The time with dual revision is:

$$\text{dual review} = T_0 + \frac{S-S_0}{v}$$

However, if the two reviewers work in different set of studies (no dual review), the time is:

$$\text{no dual review} = T_0 + \frac{S-S_0}{2v}$$

Time saving based on the number studies, S, to be analyzed is defined as:

$$ts(S) = 1 - \frac{no\ dual\ review}{dual\ review}; \qquad ts(S) = 1 - \frac{T_0 + \frac{S-S_0}{2v}}{T_0 + \frac{S-S_0}{v}}$$

$$ts(S) = \begin{cases} \frac{1}{2} - \frac{vT_0}{2S}, & \text{si } S \geq S_0 \\ 0, & \text{si } S < S_0 \end{cases}$$

Thus, if there are no extra studies to $S_0$ to be reviewed, time saving is 0. However, if more studies have to be reviewed, the function is strictly increasing with an asymptote at ½.

$$\lim_{S \to \infty} ts(S) = \frac{1}{2}$$

Figure 4 shows graphically this function (studies is on x-axis, time saving is on y-axis, v= 1, and different values for $T_0 = (5, 5/2, 5/3, ..., 5/6)$).

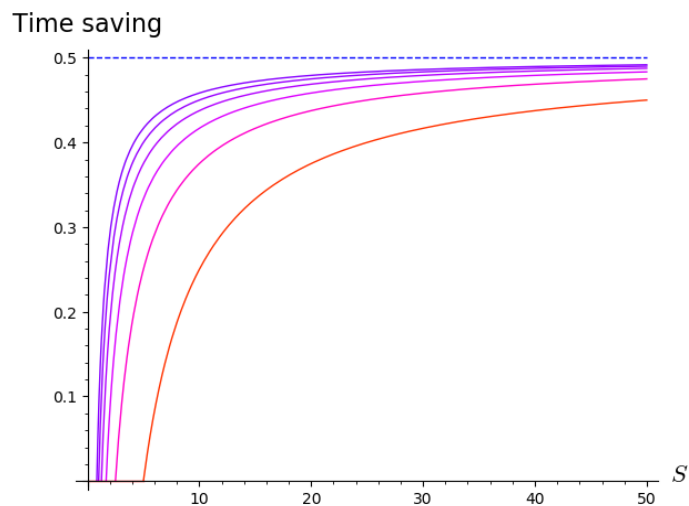

Fig. 4. Search process results